# Self-Organized Networks and Lattice Effects in High Temperature Superconductors

J. C. Phillips

Dept. of Physics and Astronomy, Rutgers University, Piscataway, N. J., 08854-8019

**Abstract**

The entirely orbital self-organized dopant percolative filamentary model describes many counter-intuitive chemical trends in oxide superconductors quantitatively, especially the high superconductive transition temperatures $T_c$. According to rules previously used successfully for network glasses, the host networks are marginally stable mechanically, and the high $T_c$'s are caused by network softening, which produces large electron-phonon interactions at interlayer dopants for states near the Fermi energy.

Scanning tunneling microscopy (STM) has revealed a strongly disordered, patchy (~ 3 nm) pattern of gap inhomogeneities in $Bi_2Sr_2CaCu_2O_{8+x}$ (BSCCO) [1-3]; because of similarities observed in angle-resolved photoemission (ARPES) on other cuprate high temperature superconductors (HTSC), this patchy pattern is probably generic. The smaller (larger) gaps are sharper (broader), and the gap average decreases with increasing doping across the superconductive phase intermediate between the insulating and Fermi liquid phase. The larger gaps persist at temperatures $T = T_p > T_c$ in underdoped samples, and are called pseudogaps [1,2]. The latter are related to "checkerboard" charge density waves (CDW) [4]. All gaps exhibit d-wave anisotropy, being largest parallel to the (10) Cu-O bonds of the $CuO_2$ planes, and zero along (11) directions. Increasing T by 14% across $T_c$ produced no measurable change in the residual pseudogap distribution in an optimally doped BSCCO sample, but greatly enhanced $N(E_F)$ [5].

The complex, strongly disordered structure of the cuprates appears to be essential to their unparalleled properties as HTSC. Experience with exponentially complex molecular glasses has shown that no single polynomial mechanism describable by mean field theory can provide a satisfactory explanation for such extreme properties; instead, one attempts to identify multiple factors, all of which are optimized. Meanwhile, "rigorous" [adiabatic crystalline] formal polynomial models leave open the question of the microscopic mechanisms responsible for the gaps, their inhomogeneities, and the origin of HTSC itself [6].

Given the complexity of strongly disordered systems, one can turn to the topological model of strongly disordered space-filling, self-organized networks [7] that has achieved wide successes in dealing with the phase diagrams of network glasses as well as HTSC [8,9]. Such networks generically exhibit *three* phases [10] in {molecular} (electronic) glasses: underconnected {soft or floppy}(insulating) (white), optimally connected {isostatic, rigid but unstressed} (strange metal) (grey), and overconnected {stressed rigid} (Fermi liquid) (black). This model of HTSC emphasizes dopant-centered coherent grey filaments that *percolate optimally* (if necessary, by tunneling) through white-and-black mazes formed by defects and either insulating or Fermi liquid nanodomains. It takes account of the host lattice through the d-wave planar anisotropy of phonon scattering (largest in (10) and smallest in (11) directions). One can interpret the d-wave anisotropies of the superconductive gap $\Delta_s$ and the pseudogap $\Delta_p$ as incidentally reflecting a local gap along each filament that varies with the angle of the tangent to the filamentary path; the local gap is formed in response to coherent phonon interactions.

The first major difference between the topological model and currently popular polynomial models based on d-wave Hamiltonians for the joint superconductive and pseudogaps [11] is that the exponentially complex topological model recognizes that the superconductive gap $\Delta_s$ and pseudogap $\Delta_p$ observed experimentally refer to *disjoint* channels in T = 0 electronic Hilbert spaces: $\Delta_s$ refers to BCS-correlated Cooper *pair* products, whereas $\Delta_p$ refers to a self-consistent CDW adequately described by a Hartree product of *one-electron* wave functions. This difference has an important effect on the canonical phase diagram, which shows $T_{cs}(z) = T_{s0}(1-z^2)$ and $T_p(z) = T_{p0}(1-z)$ [$z < 1$], where $z = x/x_0$, and $x$ ($x_0$) is the (optimal) dopant concentration. If $T_{cs} \sim \Delta_s(100)$ and $T_p(z) \sim \Delta_p(100)$, and the two gaps belonged to the same channel in Hilbert space, $T_{cs}(z)$ and $T_p(z)$ would have to exhibit some kind of coupled 2x2 matrix duality in their composition dependence (for example, both diagonal terms linear, or both quadratic, in $z$), which is certainly not the case. In the topological model the two gaps belong to different channels in Hilbert space, which are coupled by phonon absorption for $T > 0$ as the phonons distort the optimally percolative filaments: this variational explanation is characteristically topological, not analytical, in nature. It is fully consistent with the observed functional differences between $T_{cs}(z)$ and $T_p(z)$, while retaining proximity effects as the gap values are coupled through non-local strain fields and self-organized dopant configurations [12].

Plots of the planar resistivity $\rho_{ab}(z,T)$ exhibit *two* Ando lines $d^2\rho(z)/dT^2 = 0$ [13]. The first line is diagonal and defines $T_p(z)$, while the second defines a critical crossover very nearly fixed vertically at $z = 1$. The observed non-crossing of $T_c(z)$ and $T_p(z)$ is caused by $T > 0$ electron-phonon mixing of the single-particle and Cooper pair channels. The *rectilinear* (not curvilinear) nature of both Ando lines is strong evidence for the percolative character of the (super)conductive paths formed by the internal filamentary network, as such rectilinearity is characteristic of percolative phenomena in strongly disordered (glassy) systems [14]. Specifically, the flexible, hierarchically self-organized nature of this network explains [9] the step functions at $z = 1$ observed in the ARPES quasi-particle ratios $Z_{(100)}(E_F)/Z_{(110)}(E_F)$ [15] and the picosecond isoenergetic pump/probe relaxation time $\tau$ observed following femtosecond 1.5 eV pulses [16]. Of course, step functions are intrinsically *non-analytic* (not polynomially complete, NPC), and their ubiquitous appearance at $z = 1$ shows that the exponentially complex topological model (independent of T) should be the foundation of any complete theory of HTSC [6].

This reasoning can be extended to discuss the Uemura phenomenology [17] connecting $T_{cs}^{max}$ (optimal doping) and $n_s$, where $n_s$ is estimated from magnetic field penetration depths λ measured by muon spin relaxation. The data show that $T_{cs}^{max}$ is nearly linear in $n_s$ for many cuprates, including samples with $T_c$ depressed by Zn doping. Within the filamentary model $T_c^{max}$ can be estimated as follows. Each filament binds its own set of dopant-derived one-electron states that are phase-correlated to produce maximum conductivity, and hence maximum screening of fluctuating internal ionic fields, with the dopants occupying optimized curvilinear threading positions during sample annealing, and refining these positions with decreasing temperature. Below $T_c^{max}$ the fraction of filaments with mutually correlated phases is proportional to $n_s(T)/n_s(0)$. At $T = T_c^{max}$ Cooper pair phase coherence is erased by interfilamentary phonon absorption. The average spacing between planar filaments (or by three-dimensional filaments projected onto metallic planes) is d, and $dn_s \sim 1$. Thus as $n_s$ decreases, the spacing between paired filaments increases, and $T_c^{max}$ decreases with the absorbed phonon energy.

We now look for the phonon that will be most effective in displacing paired filaments from their optimal configuration. The filaments zigzag from grey dopants outside $CuO_2$ planes (where electron-phonon interactions are large, and $\Delta_s$ is large locally) through paths in black $CuO_2$ planes (where $\Delta_s$ is small and probably would be 0 except for proximity effects) to the next dopant. The parts of the filamentary path most easily disrupted are therefore the weak links in the $CuO_2$ planes, which are common to all the cuprates, and the phonon we are looking for belongs to these planes. Because the $CuO_2$ planes are the stiffest and least disordered element in the host lattice, the dispersion of these phonons is easily measured [18]: the maximum energy at the (100) longitudinal acoustic phonon at the zone boundary is $\omega_0 = 10$ meV. The actual energy of the phase-breaking phonon should be of order $T_c^{max}$. Combining these equations, one finds $T_c^{max} \sim pn_s(0)$. Thus (100) LA phonons set the overall energy scale for HTSC. The proportionality factor p is not easily estimated, as so little is known about filamentary geometry, but it is approximately independent of doping. In particular, even in underdoped samples, the filaments are broadened by phonon-induced proximity effects approaching optimal doping as the interfilamentary barrier $\Delta_p - \Delta_s \rightarrow 0$ with $z \rightarrow 1$. It is this "non-crossing" broadening due to phonons in the thermal bath that makes $T_s(z)$ flatten and appear to be quadratic while $T_p(z)$ remains linear as z increases to 1. On the overdoped side the filaments overlap to form Fermi liquid patches whose area increases as z increases above 1. We can safely assume that p is larger than in conventional superconductors because filamentary (one-dimensional) glassy topology takes advantage of self-organization to be more efficient in constructing high-conductivity vortex loops to expel or screen magnetic fields than (three-dimensional) electron gases.

The complexity of strong glassy disorder generally prevents the successful construction of polynomial models of glasses. In their place one usually finds several trends: these trends reflect the combined effects of optimization of properties of most interest, which in HTSC has consisted largely of maximizing $T_c$ (although other properties are also likely to be important for applications). There are two other trends apart from the Uemura correlation, both referring to variations in $T_c^{max}$ with host lattice properties. Both of these trends are connected with host lattice instabilities and softening; these are a characteristic feature of strong electron-phonon coupling, and have already been observed to limit $T_c^{max}$ in alloys of the old intermetallic superconductors involving (for example) NbN and $Nb_3Sn$ [19].

First-principles calculations of $T_c$ based on electron-phonon interactions (EPI) in self-consistent electronic structures with ideal atomic positions are usually quite accurate for "old" superconductors (such as $MgB_2$ [20]), but such calculations for cuprates yield $T_c$'s too small by factors ~ 100 [21]. This failure indicates the breakdown of either conventional EPI, or the ideal lattice structure, leading to enormously enhanced interactions due to the glassy character of dopant configurations; experiment has amply demonstrated that the latter violently disturb cuprate vibrational spectra [22]. Conventional lattice dynamics (even empirical spring constant models) encounters many technical difficulties in the cuprates. Not only is the number of atoms/(unit cell) large, but also the basic structural unit is actually a nanodomain containing $\sim 10^3$ vibrational degrees of freedom. Thus statistical factors become important, and because of dopant disorder the relevant statistics are those of glasses, not gases, liquids or crystals. The cuprates are closely related to peroxides (such as $BaTiO_3$), many of which are ferroelectrics, and nearly all of which are marginally stable elastically and strongly disordered, with nanodomains similar to the cuprates (for instance, manganites [23]).

Lattice softening can be calibrated by treating cuprates and perovskites as incipient glasses, subject to the same axiomatic rigidity rules as network glasses [7]. These rules are simplest and most easily justified for chalcogenide glass alloys composed of atoms of similar size [24], but more general rules have succeeded for oxide glasses, notably window glass [25], which is 74% $SiO_2$ alloyed with 16% $Na_2O$ and 10% $CaO$. These chemical proportions of window glass, an ideal, globally and locally stress-free network, are partially explained in terms of the average number $<R>$ of *Pauling* resonating valence bonds/atom, with $<R> = 2.40$ *exactly* [$R(Na) = 1$, $R(Ca) = 2$, $R(Si) = 4$, and $R(O) = 2$]. The results [24] for many ideally stress-free binary and ternary chalcogenide alloy network glasses range from $<R> = 2.27$ to $<R> = 2.52$; the entire set of ranges is centered on $<R> = 2.40$.

Given this background, is there a way to understand both ferroelectrics and cuprates? There is [26]: $<R>$ of many ferroelectric perovskites (such as $BaTiO_3$) is 2.40; these perovskites have large energy gaps, and they can be alloyed with isovalent elements ([Pb,Zr] $TiO_3$), but not doped. By decreasing $<R>$ we reach the dopably metallic HTSC cuprates, which span the range from $<R> = 1.67$ up to 2.24 (Fig. 1), which lies just below the range [2.27,2.52] of stress-free network glasses. In this range oxide crystalline networks are soft, and any metallic states at the Fermi energy should be erased by Jahn-Teller distortions, in the cuprates specifically by buckling of the tetragonal basal planes. Experimentally it has been observed that such buckling is incipient and does limit $T_c$, but the distortions are small because of the isostatic (rigid but unstressed) nature of the $CuO_2$ planes [26]. Thus these rigid planes are not the site of the strong interactions which produce HTSC: quite the opposite, those interactions occur at the dopants in soft planes outside the $CuO_2$ planes, while the planes function in two other ways: (1) as mechanical stabilizers against Jahn-Teller distortions, and (2) as electrical media through which Cooper pairs formed by strong electron-phonon interactions at dopants can connect through S(dopant)-N($CuO_2$ plane)-S(dopant) [SNS] tunneling. Note, by the way, that the oxi-chloride, $Na_xCa_{2-x-y}CuO_2Cl_2$ (NCCOC, $T_c(x = 0.2, y = 0) = 28K$, $<R(0.2,0)> = 1.67$ ) forms a poor network (because $R(Cl) = 1$), but with the assistance of Na to bridge the $CuO_2$ planes, self-organization and stabilization by a 4x4 CDW checkerboard [27], it still manages to be superconductive; replacement of Na by Ca vacancies increases $<R>$, reduces defects, and gives $T_c(x = 0, y = 0.2) = 38K$ [28]. Finally, although polaronic effects are strong in all the cuprates [29], they are especially strong in $Na_xCa_{2-x-y}CuO_2Cl_2$ because two ions (Na and Cl) have $R = 1$.

Perhaps the strongest argument against EPI as the source of HTSC has been the disappearance of the oxygen isotope shift, which is large and normal near the metal-insulator transition (MIT), but decreases towards a small (but still non-zero) value near $z = 1$ [30]. This long-standing mystery is mitigated by recognizing the variational nature of flexible self-organized percolation. Near the MIT, the paths are far apart and isotopic substitution does not alter the phonon dynamics that causes the normal isotope shift. However, at $z = 1$ the dynamic effects are compensated by the combined effects of zero-point vibrations [19] and space-filling. Site-selective isotope shifts in the host lattice [30] are an acid test for this explanation. The $CuO_2$ planes are isostatic [26] and nearly ideally crystalline, and hence exhibit an O isotope effect, but the low R planes between them (such as BaO, $R = 2$) are soft and glassy, so there is no isotope effect at the apical oxygensor the $CuO_x$ chains. This counter-intuitive result is similar to the counter-intuitive clamping (freeing) of the states between $E_F$ and $E_F + \theta_D$ (below $E_F + \theta_D$) observed by ARPES [31], and explained by glassy constraint theory [8].

A third factor involving soft lattices is the local topology associated with the (100) longitudinal optic (LO) phonon kinks [22,32]; these occur near $\mathbf{G}/4 = (1/2\,0\,0)$ and may be related to the 4x4 checkerboard pattern that appears to be associated with pseudogaps in underdoped patches [4]. Probably the most instructive data on the LO phonon anomaly are those [33] taken for YBCO at light doping (x = 0.2) below the MIT, at the MIT (x = 0.35), in the 60K plateau (x = 0.6), and at optimal doping (x= 0.92). Before long CuO chains have formed (x = 0.2 only strong LO scattering (labeled $N_2$) occurs near $\mathbf{q} = (q\,0\,0)$ with q = 3.0 at 73 meV, reflecting the strong disorder and the validity of mean-field models. As soon as the CuO chains percolate (x = 0.35), a new LO band (labeled $Z_3$) appears near the zone boundary (q = 3.5) at 57 meV. When the minor cross-linking chains have begun to percolate (x = 0.6), both $N_2$ and $Z_3$ broaden, and the gap $\Delta\omega_{LO}$ near q = 3.25 increases, indicating phase separation between the nanodomains [7] with and without minor chains. At optimal doping (x = 0.92) there is only one phase, but it is ideally glassy [8], and the mean-field component $N_2$ has become very weak, while the percolative component $Z_3$ is very strong.

As all factors must conspire variationally to produce HTSC, one can assume that the LO phonon gap $\Delta\omega_{LO}$ is one of them, and plot $T_c^{max}$ against $\Delta\omega_{LO}$ (Fig. 1(b)): again, there is a strong correlation. Considering that $T_c^{max}$ is limited by $T_p$, and that the pseudogap phase is apparently stabilized by the 4x4 checkerboard pattern associated with it, this correlation is natural. Of course, space filling produces the sharpest gap, which occurs at optimal doping ($T_c = T_c^{max}$) in $La_{2-x}Sr_xCuO_4$ (LSCO), x = 0.15; this gap is just as sharp as in the x = 1/8 crystalline "stripe" phase [32], where $T_c = 0$.

One more factor should be considered, and that is the dopant sites themselves. In the presence of multiple phases, there may be multiple dopant sites for the same dopant, one for each phase. The natural candidate for superconductive interstitial O in BSCCO in a superconductive region is a split apical (Cu-O-Bi) site, which apparently generates polaronic structure in the midinfrared [34]; this site could be associated with an impurity band pinned to $E_F$ [35]. The pseudogap regions dominate I-V STM characteristics near $E_F$ -0.9 V, and there a dopant site is found near Sr [35] that can be assigned to pseudogap regions. If there are two dopant sites, one can also consider the probable scenario in which both kinds of sites form separate and independent sets of filaments. Note that while the superconductive dopants must be metallically connected, CDW formation energies are enhanced in one dimension (which may be curvilinear), so that there may be two interpenetrating self-organized networks in the cuprates. Evidence that there is a second dopant site that generates an electrically active impurity band is apparent in the dramatic increase in $N(E_F)$ [5] as T increases across $T_c$.

The successful phenomenological correlation between $T_c^{max}$ and <R> in Fig. 1(a) is comparable to that obtained [36] between $T_c^{max}$ and $t_2/t_1$, where $t_j$ is a tight-binding overlap integral used in fitting first-principles band structures of states near $E_F$. It is not surprising that two apparently complementary microscopic variables give equally good fits; this is common in (exponentially complex) glassy contexts. Both variables have their limitations: $t_2/t_1$ describes the metallic character of the $CuO_2$ plane, and would be largest for a free electron gas, whereas <R> can achieve its optimal value in an insulator like Se, so both factors are needed to describe the microscopic origins of HTSC.

In conclusion, the rapidly expanding and improving data base for HTSC supports EPI as the dominant mechanism determining the properties of both superconductive and pseudogap phases [37]; several soft-lattice-based factors are important in describing generic trends in $T_c$ for different materials. All efforts to construct "complete" theories of HTSC [6] should recognize the importance of EPI and the exponential complexity implied by the glassy nature of these soft lattices.

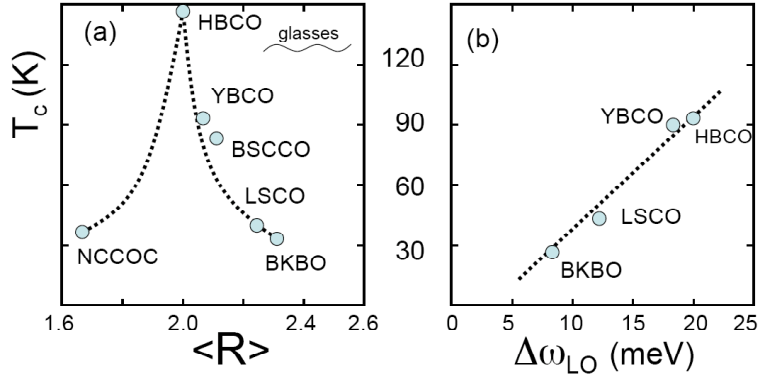

Fig. 1. (a) Chemical trends in $T_c^{max}$ with $<R>$, which measures the global stiffness of the doped crystalline network, with $R(Cu) = 2$ and $R(Bi) = 3$. Perovskites and pseudoperovskites are only marginally stable mechanically [20], and $<R>$ lies in the region of floppy networks just below the isostatic (rigid but unstressed) range determined by studies of network glasses (wavy line). The peak in $T_c^{max}$ occurs at $<R> = 2$, as one would expect from mean field percolation theory. (b) The cuprates are stabilized by checkerboard reconstruction, the strength of which may determine the $\Delta\omega_{LO}$ phonon ("half-breathing mode") anomaly, which also correlates well with $T_c$. (The single crystal sample [32] of Hg1201 had $T_c = 94K$.) Dashed lines are guides only.